\documentclass[12pt]{article}
\usepackage{amssymb,amsmath}

\newcommand{\IR}{\mathbb{R}}
\def\be{\begin{equation}}
\def\bea{\begin{eqnarray}}
\def\ee{\end{equation}}
\def\eea{\end{eqnarray}}
\def\nn{\nonumber \\}

\begin{document}
\begin{titlepage}

\begin{flushright}
IPhT-T09/020\\
DFPD09/TH05
\end{flushright}

\bigskip
\bigskip
\centerline{\Large \bf Non-BPS Black Rings and Black Holes in Taub-NUT}
\bigskip
\bigskip
\centerline{{\bf Iosif Bena$^1$, Gianguido Dall'Agata$^2$, Stefano Giusto$^1$,}}
\centerline{{\bf Cl\'{e}ment Ruef$^{\, 1}$ and Nicholas P. Warner$^3$}}
\bigskip
\centerline{$^1$ Institut de Physique Th\'eorique, }
\centerline{CEA Saclay, 91191 Gif sur Yvette, France}
\bigskip
\centerline{$^2$ Dipartimento di Fisica ``Galileo Galilei'' \& INFN, Sezione di Padova,}
\centerline{Universit\`a di Padova, Via Marzolo 8, 35131 Padova, Italy}
\bigskip
\centerline{$^3$ Department of Physics and Astronomy}
\centerline{University of Southern California} \centerline{Los
Angeles, CA 90089, USA}
\bigskip
\centerline{{\rm iosif.bena@cea.fr,~stefano.giusto@cea.fr,~clement.ruef@cea.fr, } }
\centerline{{\rm gianguido.dallagata@pd.infn.it, warner@usc.edu} }
\bigskip
\bigskip

\begin{abstract}

  We solve the recently-proposed equations describing non-BPS extremal
  multi-center configurations, and construct explicit solutions
  describing non-supersymmetric extremal black rings in Taub-NUT, as
  well as the seed solution for the most general extremal non-BPS
  under-rotating black hole in four dimensions.  We also find
  solutions that contain both a black hole and a black ring, which
  descend to four-dimensional extremal non-BPS two-center black holes
  with generic charges.

\end{abstract}

\end{titlepage}

\section{Introduction}

Supersymmetric solutions that preserve the same supersymmetries as
three-charge black holes or black rings in five dimensions are well
understood and can be written in terms of three self-dual two-forms
describing magnetic fluxes on a hyper-K\"{a}hler four-dimensional
base, three warp factors, sourced either by the two-forms or by
singular sources, and an angular momentum one-form \cite{Bena:2004de}.
These  solutions to M-theory, or type II string theory, can be recast  in terms of
BPS solutions of five-dimensional $U(1)^3$ ungauged supergravity and can also 
be easily generalized to $U(1)^N$ supergravities \cite{Gutowski:2004yv}. If the
four-dimensional hyper-K\"{a}hler base space is Gibbons-Hawking (or
Taub-NUT), the two-forms, the warp factors and the angular momentum
can be determined entirely in terms of eight ($2N+2$) harmonic
functions \cite{Gauntlett:2002nw,Gauntlett:2004qy,Bena:2005ni}, and
descend to four-dimensional BPS multi-centered black hole
configurations \cite{denef}.

Implicit in the construction of the supersymmetric solutions is the
choice of an orientation for the hyper-K\"{a}hler four-dimensional
base: The curvature tensor can be arranged to be either self-dual or
anti-self dual.  For supersymmetry it is crucial that the Riemann
curvature of this base has the same duality as the three magnetic
two-forms: They must all be self-dual or anti-self-dual.  The
difference in choice merely amounts to an overall reversal of
orientation and is usually neglected.  However, there has been a very
nice recent observation \cite{Goldstein} that one can obtain extremal
non-supersymmetric solutions of the supergravity equations of motion
by flipping the relative dualities of the hyper-K\"{a}hler base and
the magnetic two-forms\footnote{We will consistently fix our hyper-K\"{a}hler
base to be self-dual ({\it i.e.} with self-dual curvature) and so this
new prescription amounts to starting with anti-self-dual magnetic
two-forms and solving the supersymmetric BPS equations with flipped
dualities.}. This means that supersymmetries are ``locally preserved''
by the sources but globally broken by the incompatible holonomy of the
background metric on the base.

A simple example of this, and a very useful tool in our analysis, is
to start by noting that there are two ways of writing the flat metric
on $\IR^4$ in Gibbons-Hawking (GH) form: One that looks self-dual and
one that looks anti-self-dual.  While this distinction is a coordinate
artifact for $\IR^4$ (because the curvature is trivial), one of the
choices will break supersymmetry in more general backgrounds. Indeed,
it is fairly straightforward to adapt what appears as an orientation
reversal in $\IR^4$ to a highly non-trivial, supersymmetry-breaking
transformation in Taub-NUT.  Thus, given an asymptotically $\IR^4$
solution, one can find two ways of extending it to an asymptotically
Taub-NUT solution: one that preserves the supersymmetry and one that
does not.

The basic technique is also easily understood in terms of the
underlying brane construction.  For example, an asymptotically
five-dimensional black ring solution (with a flat $\IR^4$ base)
preserves the four supersymmetries respected by its three constituent
electric M2 branes.  When one replaces the $\IR^4$ base by a Taub-NUT
space and considers the solution from the IIA perspective, the M2
branes descend to D2 branes while the tip of Taub-NUT descends to a D6
brane.  In the BPS embedding, the four Killing spinors preserved by
the three sets of D2 branes are the same as those of the D6 brane, and
thus the solution is supersymmetric. In the non-BPS embedding the D6
brane has opposite orientation, and hence it does not preserve any of
the four Killing spinors of the D2 branes.

An interesting corollary of this D-brane picture  is
that five-dimensional objects that preserve the same eight Killing
spinors as two sets of M2 branes, will still be supersymmetric when
embedded in self-dual or anti-self-dual Taub-NUT. Indeed, if
only two sets of D2 branes are present, the D6 brane will be mutually
BPS with them irrespective of its orientation. Hence, a two-charge
supertube embedded in Taub-NUT in the ``duality-matched'' embedding
\cite{BenaKrausKKM} or in the ``duality-flipped'' embedding \cite{Goldstein} will still be
supersymmetric. We will see in Section 4 the rather unexpected
fashion in which this is realized.

Our purpose in this paper is to give a general algorithm for
constructing the most general two-center solution of the ``almost BPS
equations'' presented in \cite{Goldstein}. The most obvious solution
to look for is a ``non-BPS'' two-charge supertube in Taub-NUT. However, as
we explained above, this solution turns out to be identical to that of
the BPS supertube in Taub-NUT. The next obvious solution is the
non-BPS three-charge three-dipole charge black ring in Taub-NUT, which
we construct in Section 3.

Because the new non-BPS black-ring solution becomes identical to the
BPS solution both in $\IR^4$, and in $\IR^3 \times S^1$, it is
possible to recycle many of the pieces of the BPS three-charge
three-dipole charge black ring solutions in $\IR^3 \times S^1$ and
$\IR^4$
\cite{Bena:2004wv,Elvang:2004rt,Elvang:2004ds,Bena:2004de,Gauntlett:2004qy},
and the only new ingredient is to solve one non-trivial equation for a
piece of the rotation vector.  The full solution is again generated from 
several harmonic functions, determining the M2 charges, M5 dipole
charges and angular momentum of the black ring. However, these
harmonic functions enter the solution very differently than for BPS
black rings in Taub-NUT
\cite{Elvang:2005sa,Gaiotto:2005xt,Bena:2005ni}. Furthermore, in order
for the solution to be free of closed timelike curves (CTC's), the
harmonic function that determines the angular momentum must have both
a $1/r$ source at the tip of Taub-NUT, as well as a ``dipole'' piece
of the form $\cos \theta /r^2$ centered at the black ring location.
No such terms appear in the BPS ring solution, and the necessity of
their presence is far from obvious without a careful construction of
the full solution.

Since these new solutions ``locally preserve'' supersymmetry but break
it globally, one expects that local properties should be the same as
those of the BPS counterparts.  Indeed, we find that the near-horizon
geometry of the non-BPS extremal ring is identical to that of its BPS
cousin, and its entropy is given by the $E_{7(7)}$ quartic invariant
as a function of its charges \cite{Bena:2004tk}. On the other hand, the location, or
``radius'' of the ring in Taub-NUT is a more global property and is
generically different for BPS and non-BPS solutions.  For both BPS and
non-BPS solutions the location is determined by the requirement
that there be no Dirac-Misner strings, but the source terms that can
give rise to such strings are very different for BPS and non-BPS
solutions.  We also show, in Section 4, that when black rings are
reduced to two-charge supertubes, the BPS and non-BPS solutions
coincide, and the two radii become equal.

As observed in \cite{Goldstein}, the almost BPS equations can be used
to re-derive the non-rotating extremal non-BPS four-dimensional
single-center black hole obtained in
\cite{LopesCardoso:2007ky,Gimon:2007mh}. However their power is much
greater, even for single-center solutions: by adding to the angular
momentum harmonic function a ``dipole'' piece of the form $\cos \theta
/r^2$ centered at the black hole location, we can give this black hole
rotation. The resulting solution is a new rotating extremal non-BPS
solution in four dimensions. This solution has five (four-dimensional)
quantized charges (corresponding to D6, D0 and three sets of D2
branes) as well as angular momentum\footnote{It is also
trivial to introduce Wilson lines for the magnetic gauge fields,
because they do not affect the rest of the solution in any way (unlike
for BPS solutions).}.

For particular values of the charges and moduli one can show that this
black hole can be related by dualities to the ``slowly-rotating'' or
``ergo-free'' extremal limit\footnote{See, for example,
  \cite{Emparan:2007en} or \cite{Astefanesei:2006dd} for a discussion
  of the two extremal limits of this black hole.} of the D6-D0
(Rasheed-Larsen) black hole \cite{rasheed-larsen} or its D6-D2-D2-D0
dual \cite{Giusto:2007tt}.  However, our solution is much more
general, as it can have arbitrary D6-D2-D2-D2-D0 charges.
Hence this solution is the seed solution for the most generic extremal
under-rotating black hole of the $STU$ model and of ${\cal N}=8$
supergravity in four dimensions.

Using our method it also is quite straightforward to find a solution
that contains both this generic rotating black hole and a black ring.
The presence of the black hole adds an extra source term to the black
ring warp factor, and three more terms to the angular momentum vector.
It also modifies the black ring radius relation, without changing the
near-horizon geometry of either the ring or the hole.

The non-BPS black ring with a black hole in the middle can be
compactified to four dimensions, to give a two-center non-BPS
solution, with a non-trivial angular momentum. The black hole at one
of the centers has five charges (D6-D2-D2-D2-D0), and the black hole
at the other center has seven charges (D4-D4-D4-D2-D2-D2-D0). Since we
find the solution for arbitrary moduli, this system can be dualized
into one where each of the two black holes has D6-D4-D2-D0 charges
and can probably be identified to the most generic extremal two-centered
solution of the $STU$ model.

Before beginning, it is important to note that there exists a rather
large body of work on constructing extremal black holes in
four-dimensional supergravity, that started from the observation of
\cite{Ceresole:2007wx} that the second-order equations underlying
these solutions can be factorized as products of easier-to-solve
first-order equations\footnote{See, for example,
  \cite{LopesCardoso:2007ky,Andrianopoli:2007gt,Gimon:2007mh,Ferrara:2008ap,Bellucci:2008sv}.}.
So far, the single-center solutions obtained in this way appear to be
captured in the ansatz in \cite{Goldstein}. On the other hand there
exists a rather complementary body of work on embedding non-extremal
five-dimensional solutions in Taub-NUT, that began with
\cite{Ford:2007th,Giusto:2007fx} and resulted in the recent
construction of non-extremal black rings in Taub-NUT
\cite{Camps:2008hb}. It would be interesting to see if one can
construct our extremal non-BPS ring using either of these approaches,
and whether, upon extending these approaches to construct our
solution, one could access to a larger set of solutions than those
contained in the ansatz of \cite{Goldstein}.

In Section 2 we review the ansatz of \cite{Goldstein} for finding
non-BPS solutions. In Section 3 we outline our solution-finding
technique by constructing a three-charge three-dipole
non-BPS black ring in Taub-NUT. We also analyze its charges, mass and
near-horizon limit. In Section 4 we construct a non-BPS supertube in
Taub-NUT, and show that this is identical to a BPS supertube. In
Section 5 we construct a five-charge rotating black hole, which is the
seed solution for the most general extremal non-BPS
under-rotating black hole in four dimensions. We also discuss its relation
to the Rasheed-Larsen solution. In Section 6 we construct a solution
that includes both a rotating black hole at the tip of Taub NUT and a
black ring; this solution descends to a two-centered non-BPS black
hole solution in four dimensions. We conclude in Section 7.

\section{``Almost BPS'' solutions}

BPS solutions of eleven-dimensional supergravity carrying M2 and M5 charges are of the form
\bea
ds^2&\!\!\!\!=\!\!\!\!& -(Z_1 Z_2 Z_3 )^{-2/ 3}(dt+k)^2 + (Z_1 Z_2 Z_3)^{1/3} ds^2_4 \nn
&&+ \Bigl({Z_2 Z_3\over Z_1^2}\Bigr)^{1/3} (dx_1^2+dx_2^2)+\Bigl({Z_1 Z_3\over Z_2^2}\Bigr)^{1/3} (dx_3^2+dx_4^2)+\Bigl({Z_1 Z_2\over Z_3^2}\Bigr)^{1/3} (dx_5^2+dx_6^2)\label{11D}\\
C^{(3)}&\!\!\!\!=\!\!\!\!&\! \Bigl(a_1 - {dt+k\over Z_1}\Bigr)\!\!\wedge \!dx_1\!\wedge \!dx_2 +\!  \Bigl(a_2 - {dt+k\over Z_2}\Bigr)\!\!\wedge\! dx_3\!\wedge\! dx_4 +\! \Bigl(a_3 - {dt+k\over Z_3}\Bigr)\!\!\wedge\! dx_5\!\wedge\! dx_6 \,,
\eea
where $ds^2_4$ is a hyper-K\"ahler four-dimensional metric.
Defining the ``dipole'' field strengths as
\be
\Theta_I= d a_I\,\,,\,\, I =1,2,3\,,
\ee
the equations following from supersymmetry for a self-dual  hyper-K\"ahler base metric are\footnote{If one uses a hyper-K\"ahler base with an anti-self-dual  curvature then the dualities in (\ref{BPSeqnsa})--(\ref{BPSeqnsc})  are flipped  to the form (\ref{thetaeq})--(\ref{keq}).}:
\bea
&&\Theta_I = *_4 \Theta_I \,, \label{BPSeqnsa}\\
&&d*_4d Z_I = {|\epsilon_{IJK}|\over 2} \Theta_J \wedge \Theta_K\,, \label{BPSeqnsb}\\
&&dk+*_4 dk = Z_I \Theta_I\,,
\label{BPSeqnsc}
\eea
where $*_4$ is the Hodge duality operation performed with the metric
$ds^2_4$.    The foregoing equations also govern the solutions of arbitrary
$U(1)^N$ ungauged supergravities in five dimensions
\cite{Gutowski:2004yv} if one replaces the $|\epsilon_{IJK}| $ by the
corresponding triple intersection number $C_{IJK}$.

It was observed in \cite{Goldstein} that a class of extremal solutions of the equations of motion is obtained by reversing the duality of the $\Theta_I$ and of $k$ relative to the duality of the curvature of the four-dimensional  base.  That is, one preserves the metric, $ds^2_4$, and the duality of its Riemann tensor but flips $*_4\to -*_4$ in (\ref{BPSeqnsa})--(\ref{BPSeqnsc}):
 \bea
&&\Theta_I =- *_4 \Theta_I \label{thetaeq}\\
&&d*_4d Z_I = {C_{IJK}\over 2} \Theta_J \wedge \Theta_K\label{zeq}\\
&&dk-*_4 dk = Z_I \Theta_I\label{keq}\,.
\eea

When the base metric $ds^2_4$ is flat $\mathbb{R}^4$, the flip of
orientation can be re-written as a change of coordinates, and
solutions to equations (\ref{thetaeq})--(\ref{keq}) are still BPS.
When $ds^2_4$ is not flat, as in Taub-NUT space, equations
(\ref{thetaeq})--(\ref{keq}) define, in general, non-BPS solutions,
which were named ``almost BPS'' in \cite{Goldstein}.

\subsection{Gibbons-Hawking base}

As with the BPS solutions, equations (\ref{thetaeq})--(\ref{keq}) are easier to solve if one specializes to Gibbons-Hawking base metrics:
\bea
ds^2_4= V^{-1}(d\psi + \vec{A})^2 + V ds^2_3\,,\qquad *_3 d{\vec A} = d V\,.
\label{GHmet}
\eea
We will also only look for solutions that are invariant under $\psi$-translations.

The four-dimensional  geometry is encoded in the function $V$, which is harmonic with respect to the flat three-dimensional euclidean metric
 $ds^2_3$. The Hodge star operation in $\mathbb{R}^3$ is denoted by $*_3$ and one-forms on $\mathbb{R}^3$ are denoted by a vector superscript.  In general, for a GH base one can take $*_3 d{\vec A} = \pm  d V$ and this leads to self-dual or anti-self-dual Riemann tensors.
The choice in (\ref{GHmet}) means we are choosing a self-dual curvature.

The one-form potentials   for the anti-self dual field strengths have the form:
\be
a_I = K_I (d\psi + \vec{A}) + \vec{a}_I\,,\qquad *_3 d \vec{a}_I = V d K_I - K_I dV\,,
\label{dipolepotential}
\ee
where $K_I$ is a harmonic function  on $\mathbb{R}^3$. Such $a_I$'s thus provide the general solution to eq. (\ref{thetaeq}).

Using this result in eq. (\ref{zeq}), one finds that the warp factors $Z_I$
must satisfy
\be
d*_3 d Z_I = {1\over 2} \, C_{IJK} \,V d *_3 d (K_J K_K) \,.
\ee
Unlike the BPS solution, this equation does not, in general, admit a closed form
solution written solely in terms of the functions $V$ and $K_I$.  However, in practice, it is still relatively
straightforward to obtain exact solutions for $Z_I$.

Expanding $k$ along the fiber and base of the Gibbons-Hawking space:
\be
k ~=~\mu (d\psi +\vec{A})+\vec{\omega}\,,
\ee
one can reduce (\ref{keq}) to:
\be
d (V \mu) ~+~ *_3 d\vec{\omega} ~=~ V Z_I dK_I\,.
\label{redeqn}
\ee
Acting with $d*_3$  one obtains the following equation for $\mu$:
\be
d*_3 d (V \mu) ~=~  d(V Z_I) *_3 \wedge\, d K_I\,.
\ee
This equation is the integrability condition for (\ref{redeqn}).
Again, one does not seem to be able to find a simple, general solution to this equation, but we
will obtain particular solutions in later sections.

\section{Non-BPS extremal black ring}
\label{ringsec}

In this section we derive one of the main results of this paper: an exact
solution representing a non-BPS extremal regular black ring in
Taub-NUT space. This space is described by the Gibbons-Hawking potential
\be
 V = h ~+~ {Q_6\over r}\quad \Rightarrow \quad\vec{A}~=~ Q_6 \cos\theta d\phi\,.
 \ee
 We have introduced a generic constant $h$ in $V$ to facilitate comparison
 with the flat space ($\IR^4$) limit, which corresponds to taking $h=0$. Taking
 $Q_6=0$ corresponds to the infinite radius limit of the
 black ring, in which the base reduces to $\mathbb{R}^3\times S^1$. In
 both of these limits the non-BPS solution must reduce to the known BPS
 black ring solution.

\subsection{Solving the equations}

 We take the position of the black ring in $\mathbb{R}^3$ to be along
 the positive $z$ axis at a distance $R$ from the origin of Taub-NUT.
 We denote polar coordinates centered at the black ring position by
 $(\Sigma,\theta_\Sigma)$. Their relation to the polar coordinates
 $(r,\theta)$ centered at the origin is:
\be
\Sigma = \sqrt{r^2 + R^2 - 2 r R \cos\theta}\,,\qquad \cos\theta_\Sigma = {r\cos\theta-R\over \Sigma}\,.
\label{polarSigma}
\ee

The black ring carries dipole charges associated with the harmonic functions\footnote{As one can see from (\ref{dipolepotential}), adding a constant $\kappa_I$ to $K_I$ has the only effect of shifting the dipole potential $a_I$ by the constant one-form $k_I d\psi$. Hence a constant in $K_I$ is physically irrelevant.}
\be
K_I ~=~{d_I\over \Sigma}\,, \qquad I=1,2,3 \,.
\ee
According to eq. (\ref{dipolepotential}), the corresponding dipole gauge fields are given by:
\bea
a_I = {d_I\over \Sigma} (d\psi + \vec{A}) + \vec{a}_I \,,\quad \vec{a}_I=h \,d_I {r\cos\theta-R\over \Sigma} d\phi + Q_6 d_I {r-R\cos\theta\over R\Sigma} d\phi\,.
\label{dipolenonbpsring}
\eea
The warp factors $Z_I$ are determined by the equation:
\be
d*_3 d Z_I = {C_{IJK}\over 2} V d*_3 d (K_J K_K)={C_{IJK}\over 2}  \Bigl(h+{Q_6\over r}\Bigr) d*_3 d \Bigl({d_J d_K\over \Sigma^2}\Bigr)\,.
\ee
The solution $Z_I$ can be written as the linear combination of two terms. The first term satisfies the equation:
\be
d*_3 d Z_I^{(1)} ={C_{IJK}\over 2}\, h\,d*_3 d \Bigl({d_J d_K\over \Sigma^2}\Bigr)\,,
\ee
which is trivially solved by:
\be
Z_I^{(1)} = {C_{IJK}\over 2} \,h\,{d_J d_K\over \Sigma^2}\,.
\ee
The second term is found by solving
\be
d*_3 d Z_I^{(2)} ={C_{IJK}\over 2} {Q_6\over r} d*_3 d\Bigl({d_J d_K\over \Sigma^2}\Bigr)\,.
\ee
This  is the same equation as the one in a flat $\mathbb{R}^4$ base and BPS and
``almost BPS'' solutions are related by simple change of coordinates (essentially, the exchange of the coordinates $\psi$ and $\phi$).  One can therefore borrow the known BPS solution and see that the equation above is solved by:
\be
Z_I^{(2)} = {C_{IJK}\over 2} {Q_6\,d_J d_K\over R^2} {r\over \Sigma^2}\,.
\ee
Moreover we can add to $Z_I$ a harmonic function $L_I$, which has a pole at the location of the ring:
\be
L_I =l_I+ {Q_I\over \Sigma}\,.
\ee
It is not much more difficult to add a pole in $L_I$ at the center of the TN space, which corresponds to placing a black hole inside the black ring.  We will construct this more general solution in section  \ref{bhsec}.
The total solution for $Z_I$ is then
\be
Z_I = l_I+{Q_I\over \Sigma}+{C_{IJK}\over 2} {d_J d_K\over \Sigma^2} \Bigl(h+ {Q_6 r\over R^2}\Bigr)\,.
\ee

The equation for $k=\mu (d\psi+\vec{A})+\vec{\omega}$ is now:
\bea
&&d(V\mu) + *_3 d \vec{\omega} = V Z_I dK_I \label{keqring}\\
&&= \Bigl[\Bigl(h+{Q_6\over r}\Bigr)\Bigl(l_I+{Q_I\over \Sigma}\Bigr) + \Bigl(h^2+{Q_6^2\over R^2} + Q_6 h \Bigl({1\over r}+{r\over R^2}\Bigr)\Bigr){C_{IJK}\over 2} {d_J d_K\over \Sigma^2}\Bigr]d\Bigl({d_I\over \Sigma}\Bigr)\,, \nonumber
\eea
and we then expand the source term on the right-hand side into simpler component pieces. It is then straightforward  to find a solution for each piece. We list in the following the solutions for the various terms:
 \bea
&&d(V\mu_1) + *_3 d \vec{\omega_1}= \Bigl(h+{Q_6\over r}\Bigr) l_I d\Bigl({d_I\over \Sigma}\Bigr)\\
&&\,\,\Rightarrow\,\,
\mu_1 = {l_I d_I\over 2 \Sigma}\,,\quad \vec{\omega}_1 = {h\,l_I d_I\over 2}{r\cos\theta-R\over \Sigma}d\phi +{Q_6 l_I d_I\over 2}{r-R\cos\theta\over R \Sigma} d\phi\,.\nonumber
\eea
\bea
d(V\mu_2) + *_3 d \vec{\omega_2}=h\,{Q_I\over \Sigma}d\Bigl({d_I\over \Sigma}\Bigr)\quad \Rightarrow \quad
\mu_2 = h\,{Q_I d_I\over 2V\, \Sigma^2}\,,\quad \vec{\omega_2}=0\,.
\eea
\be
d(V\mu_3) + *_3 d \vec{\omega_3}={Q_6\over r}{Q_I\over \Sigma} d\Bigl({d_I\over \Sigma}\Bigr)\,.
\ee
\be
d(V\mu_4) + *_3 d \vec{\omega}_4 = \Bigl(h^2+{Q_6^2\over R^2} \Bigr) {C_{IJK}\over 2} {d_J d_K\over \Sigma^2}\,d\Bigl({d_I\over \Sigma}\Bigl)\,.
\ee
\be
d(V\mu_5) + *_3 d \vec{\omega}_5 =Q_6 h \Bigl({1\over r}+{r\over R^2}\Bigr){C_{IJK}\over 2} {d_J d_K\over \Sigma^2} d\Bigl({d_I\over \Sigma}\Bigr)\,.
\label{mu5eq}
\ee

To find a solution to the third equation it is useful to reinterpret it as the equation for a one-form
$\tilde k\equiv  r V \mu_3  (d\psi+\vec{A})+\vec{\omega_3}$ in a flat $\mathbb{R}^4$ base, and use the fact that BPS and almost BPS solutions are related by a $\psi\leftrightarrow \phi$ exchange, in flat space.  In this way one arrives at the following solutions
\be
\mu_3 =Q_6 Q_I d_I {\cos\theta\over 2 R \,V \Sigma^2}\,,\qquad \vec{\omega}_3=Q_6 Q_I d_I {r\sin^2\theta\over 2 R \,\Sigma^2}d\phi\,.
\ee

For the fourth equation one can easily verify that the following expressions
\be
\mu_4^{(1)} = \Bigl(h^2+{Q_6^2\over R^2} \Bigr) {C_{IJK}\over 6} {d_I d_J d_K\over V\, \Sigma^3}\, ,\qquad \vec{\omega_4}^{(1)}=0\,,
\ee
and
\be
\mu_4^{(2)}= \Bigl(h^2+{Q_6^2\over R^2} \Bigr) {C_{IJK}\over 6}\,d_I d_J d_K {r \cos\theta\over R V\, \Sigma^3}\,,\quad \vec{\omega_4}^{(2)}=
 \Bigl(h^2+{Q_6^2\over R^2} \Bigr){C_{IJK}\over 6}\,d_I d_J d_K {r^2 \sin^2\theta\over R\, \Sigma^3}d\phi.
\ee
both solve the equation.  Hence we will take
\be
\mu_4 = \mu_4^{(2)} + \alpha(\mu_4^{(2)}-\mu_4^{(1)})\,,\quad \vec{\omega}_4=(1+\alpha) \vec{\omega}_4^{(2)} \,,
\ee
and, for the moment, we will keep the parameter, $\alpha$, arbitrary. 

The fifth equation is the only one whose solution cannot be found by simply
recycling pieces of the black ring solutions in $\IR^4$ or $\IR^3
\times S^1$, because the right hand side vanishes in both limits ($Q_6
\rightarrow 0$ or $h \rightarrow 0$). However, it is possible to
think about the right hand side as coming from a fake solution in $\IR^4$
whose warp factor is
\be
Z_{\rm fake} ~\sim~ {r^2 + R^2 \over \Sigma^2}\,.
\ee
One can then express $Z_{\rm fake}$ in the $x$-$y$ coordinate system used
to find the black ring in $\IR^4$ \cite{Elvang:2004rt}, solve the
corresponding equations\footnote{Equations (46) and (47) in
  \cite{Bena:2004de}.} for $k_1$ and $k_2$, and express the $\IR^4$ solution
as a solution of the almost BPS equations to read off $\mu_5 V $ and $ \vec{\omega}_5$. This gives
\bea
\mu_5 &=&Q_6 h {C_{IJK}\over 6}\,d_I d_J d_K  {3 r^2 + R^2\over 2 R^2 V\,r \,\Sigma^3}\,,\\
 \vec{\omega}_5  &=& Q_6  h  {C_{IJK}\over 6}\,d_I d_J d_K  {r (3 R^2 + r^2)- R (3 r^2 + R^2) \cos\theta\over 2 R^3\, \Sigma^3} d\phi\,,
\eea
which one can also verify directly to be a solution of (\ref{mu5eq}).

Finally one has the freedom to add a solution of the homogeneous equation, that is, a one-form in TN space with self-dual field strength. Such a one-form has the general form
\be
k = {M\over V}(d\psi+\vec{A})+\vec{\omega}\,,\quad *_3 d\vec{\omega} = - dM\,,
\ee
with $M$ any harmonic form on $\mathbb{R}^3$. We take $M$ of the form
\be
M = m_0 + {m\over \Sigma}+{\tilde m\over r}\,.
\ee
We will see that, unlike the BPS solution,  a pole in $M$ at $r=0$ is necessary to produce a regular solution. Hence the final possible contributions to $\mu$ and $\vec{\omega}$ are
\be
\mu_6 = {m_0\over V}+ {m\over V\,\Sigma}+ {\tilde m\over  V\, r }\,,\qquad \vec{\omega}_6 = - m {r \cos\theta- R\over \Sigma} d\phi-{\tilde m}\cos\theta d\phi\,.
\ee

We should also note that one should think of the term proportional to
$\alpha$ in $\mu_4$ and $\vec{\omega}_4$ as coming from an extra
harmonic term in M. Thus, the harmonic function $M$ that determines
the black ring solution is really
\be
M = m_0 + {m\over \Sigma}+{\tilde m\over r} + {\alpha  {C_{IJK}\over 6 R}\,d_I d_J d_K
\Bigl(h^2+{Q_6^2\over R^2} \Bigr)}{\cos \theta_{\Sigma} \over \Sigma^2} \,,
\ee
where $ \theta_{\Sigma}$ was defined in (\ref{polarSigma}). In the
next section we will show that the coefficient of the dipole term,
${\cos \theta_{\Sigma} \over \Sigma^2}$, is fixed by requiring
regularity at the black ring horizon. We will see in Section
(\ref{bh0sec}) that such a term is not fixed by regularity at black-hole horizons, and 
in fact is required for allowing the black hole to rotate.

Adding all the terms together, we arrive at the final answer
\bea
\mu &=& {m_0\over V}+ {m\over V \,\Sigma}+ {\tilde m\over V\,r }+{l_I
  d_I\over 2 \Sigma}+{h Q_I d_I\over 2V\, \Sigma^2}+Q_6 Q_I d_I
{\cos\theta\over 2 R \,V \Sigma^2}\nn
&&+ {C_{IJK}\over 6}\,d_I d_J d_K
\Bigl[\Bigl(h^2+{Q_6^2\over R^2} \Bigr) \Bigl( {r\cos\theta\over
  R\,V\, \Sigma^3}+\alpha {r\cos\theta-R\over R\,V\, \Sigma^3}\Bigr)+
Q_6 h {3 r^2 + R^2\over 2 R^2 V\,r \,\Sigma^3}\Bigr]\,,\nn
\vec{\omega}&=&\Bigl[\kappa- m {r \cos\theta- R\over \Sigma} -{\tilde
  m} \cos\theta + {h l_I d_I\over 2}{r\cos\theta-R\over \Sigma}+{Q_6
  l_I d_I\over 2}{r-R\cos\theta\over R \Sigma} \nn
&& + Q_6 Q_I d_I
{r\sin^2\theta\over 2 R \,\Sigma^2} + \Bigl(h^2+{Q_6^2\over R^2}
\Bigr) {C_{IJK}\over 6}\,d_I d_J d_K (1+\alpha) {r^2 \sin^2\theta\over
  R\, \Sigma^3} \nn
&&+ Q_6 h {C_{IJK}\over 6}\,d_I d_J d_K {r (3 R^2 +
  r^2)- R (3 r^2 + R^2) \cos\theta\over 2 R^3\, \Sigma^3}\Bigr]
d\phi\,.
\eea
We have included a constant term $\kappa d\phi$  in $\vec{\omega}$ and this will be needed to cancel Dirac-Misner strings.

\subsection{Regularity}

The angular coordinates $\psi$ and $\phi$ both shrink to zero size at the center of Taub-NUT space, $r=0$. Hence regularity of the one-form $k$ requires that $\mu$ and $\vec{\omega}$ vanish at $r=0$ and imposes the following constraints on the parameters of the solution:
\bea
&&\mu_{r=0}=0\quad \Rightarrow \quad {\tilde m\over Q_6}+{l_I d_I\over 2 R}+ {C_{IJK}\over 6}\, {h \,d_I d_J d_K\over 2 R^3}=0\,,\label{mu0}\\
&&\vec{\omega}_{r=0}=0 \quad \Rightarrow \quad \kappa+m-{h l_I d_I\over 2}-\Bigl({\tilde m} +{Q_6 l_I d_I\over 2 R} +  {C_{IJK}\over 6}{Q_6 h \,d_I d_J d_K\over 2 R^3} \Bigl)\cos\theta=0\,.\label{omega0}
\eea
Moreover the coordinate $\phi$ degenerates on the $z$ axis ({\it i.e.} for $\theta=0$ or $\pi$): one should thus require that $\vec{\omega}$ vanishes on this axis. The constraint one obtains for
$\theta=\pi$ is
\be
\vec{\omega}_{\theta=\pi}=0\,\,\Rightarrow\,\,\kappa+m-{h l_I d_I\over 2}+\Bigl({\tilde m} +{Q_6 l_I d_I\over 2 R} + {C_{IJK}\over 6}{Q_6 h \,d_I d_J d_K\over 2 R^3}\Bigl)=0\,,
\ee
and is thus already implied by the two previous constraints (\ref{mu0}) and (\ref{omega0}). Vanishing of $\vec{\omega}$ at  $\theta=0$ imposes the further condition
\be
\vec{\omega}_{\theta=0}=0 \quad \Rightarrow \quad \kappa-{\tilde m} +\mathrm{sign}(r-R)\Bigl(-m+{h l_I d_I\over 2} +{Q_6 l_I d_I\over 2 R} +{C_{IJK}\over 6}{Q_6 h \,d_I d_J d_K\over 2 R^3}\Bigl)=0\,.\label{omegath0}
\ee
All the regularity conditions are solved by taking
\bea
m &=& \Bigl(h+{Q_6\over R}\Bigr){l_I d_I\over 2}+ {C_{IJK}\over 6}{Q_6 h \,d_I d_J d_K\over 2 R^3}\,, \nn
{\tilde m}&=& \kappa = -Q_6 \Bigl({l_I d_I\over 2 R}+ {C_{IJK}\over 6}{h \,d_I d_J d_K\over 2 R^3}\Bigr)\,.
\label{regularityring}
\eea

The parameter $\tilde m$ determines the value of $\mu$ at the center
of Taub-NUT, and the second equation determines the value of this
parameter that gives regular geometries (much like for BPS solutions).
As we will see later, the parameter $m$ gives the D0 charge of the
ring, and hence the first equation determines the distance between the
two centers, $R$, as a function of the charges. This equation is the
generalization of the bubble equations
\cite{denef,Bena:2005va,Berglund:2005vb,Saxena:2005uk} to non-BPS
black holes, and reduces to these equations in the BPS limits ($h
\rightarrow 0$ or $Q_6 \rightarrow 0$).  For BPS solutions this
equation is a simple, linear equation for $R$, but for the non-BPS
solutions this equation is cubic in $R$, and its  structure is much richer.
Since the charges of the black ring are quantized, for given values of the moduli this equation quantizes the possible values of $R$. 

Note that the foregoing conditions do not depend upon the parameter 
$\alpha$ that governs the ``dipole'' piece, proportional to ${\cos
  \theta_{\Sigma} \over \Sigma^2}$, in $\mu$.  We will see in the next
subsection that a careful analysis of regularity near the horizon
fixes $\alpha$ to a non-zero value.

We should note that the authors of \cite{Goldstein} conjectured
some expressions for the harmonic functions that underlie the non-BPS black
ring solution.  The proposed solutions for  $K_I, L_I$ and $M$ had poles at the black ring
location (much like for BPS black rings) but our analysis here
shows that such a solution will always be pathological.  Regular
solutions must have  a source in $M$ at the center of
Taub-NUT, with coefficient $\tilde m$ given by (\ref{regularityring}).  Similarly, there
must also be  very specific, non-zero ``dipole'' pieces, proportional to $\alpha$, in $\mu$ and $\vec{\omega}$.

\subsection{Near-horizon geometry}

We now examine the metric in the vicinity of the horizon, which is
located at $\Sigma=0$. We will work in the coordinates
$(\Sigma,\theta_\Sigma)$ defined in (\ref{polarSigma}). Neglecting the
torus directions $x_i$, the horizon is spanned by the coordinates
$\psi$, $\phi$ and $\theta_\Sigma$, and its induced metric (in the
eleven-dimensional Einstein frame) is
\bea
ds^2_H &=& {I_4\over (Z_1 Z_2 Z_3)^{2/3} V^2}(d\psi+\vec{A})^2 -2 {\mu \,\omega_\phi\over (Z_1 Z_2 Z_2)^{2/3}} (d\psi+\vec{A}) d\phi\nn
&&+(Z_1 Z_2 Z_3)^{1/3} \Bigl(V\Sigma^2 \sin^2\theta_\Sigma - {\omega_\phi^2 \over Z_1 Z_2 Z_3}\Bigr) d\phi^2 + (Z_1 Z_2 Z_3)^{1/3} V \Sigma^2 d\theta_\Sigma^2\,,
\label{dshorizon}
\eea
where
\be
 I_4= Z_1 Z_2 Z_3 V - \mu^2 V^2\,.
\ee
The volume element of this metric is
\be
\sqrt{g_H}=\Sigma (I_4 \Sigma^2 \sin^2\theta_\Sigma - \omega_\phi^2 )^{1/2}\,.
\ee
For generic values of the parameter $\alpha$ one has
\be
I_4\sim \Sigma^{-5}\,,\quad \omega_\phi\sim \Sigma^{-1}\,,
\ee
and thus $\sqrt{g_H}\sim \Sigma^{-1/2}$. So for generic $\alpha$ the
geometry does not have a regular horizon of finite area. However the
term of order $\Sigma^{-5}$ in $I_4$ can be canceled by taking
\be
\alpha = -{h^2 R^2\over h^2 R^2 + Q_6^2}\,.
\label{alphavalue}
\ee
One can think about $\alpha$ as the coefficient of a harmonic function
that determines a momentum one-form whose field strength is self-dual,
and hence lies in the kernel of the $(1-*)d$ operator in equation
(\ref{keq}). Adding this self-dual piece with the right coefficient is
crucial for the regularity of the solution.

For this value of $\alpha$, the metric coefficients have the following near-horizon expansions:
\bea
\label{limit1}
I_4 &=& {J_4\over \Sigma^4}+\Bigl({C_{IJK}\over 6} \hat{d}_I\hat{d}_J \hat{d}_K\Bigr)^2{Q_6^2\over R^4 \,V_R^4\, \Sigma^4}\sin^2\theta_\Sigma + O\Bigl({1\over \Sigma^3}\Bigr)\\
\label{limit2}
Z_I&=& {C_{IJK}\over 2} {\hat{d}_J \hat{d}_K\over V_R\,\Sigma^2}+ O\Bigl({1\over \Sigma}\Bigr)\\
\label{limit3}
\mu&=& {C_{IJK}\over 6} {\hat{d}_I\hat{d}_J \hat{d}_K\over V_R^2\,\Sigma^3}+ O\Bigl({1\over \Sigma^2}\Bigr)\\
\label{limit4}
\omega_\phi &=& {C_{IJK}\over 6} {Q_6^2\,\hat{d}_I\hat{d}_J \hat{d}_K\over R^2 \,V_R^2\, \Sigma}\sin^2\theta_\Sigma + O(\Sigma^0)\,,
\eea
where  $J_4$ is the usual quartic invariant:
\be
J_4(Q_I, \hat{d}_I,\hat{m}) = {1\over 2}\sum_{I<J} \hat{d}_I\hat{d}_J\,Q_I  Q_J -{1\over 4}\sum_I  \hat{d}_I^2\, Q_I^2 -{C_{IJK}\over 3}  \hat{m} \,\hat{d}_I \hat{d}_J \hat{d}_K\,.
\ee
We have also defined the ``effective'' dipole and angular momentum parameters of the ring, $\hat{d}_I$, $\hat{m}$,  via:
\be
{\hat d}_I = V_R\, d_I\,,\quad {\hat m}=V_R^{-1} \,m\,,\quad V_R = \Bigl(h+{Q_6\over R}\Bigr)\,.
\ee
One can see from these expressions that the horizon volume element has a finite limit for $\Sigma\to 0$:
\be
\sqrt{g_H}\to J_4^{1/2} \sin\theta_\Sigma\,,
\ee
and that the five-dimensional horizon area is given by
\be
A_H = (4\pi Q_6) (4\pi) J_4^{1/2}\,.
\ee

To compare this area to that of the BPS black ring in Taub-NUT, it is
easiest to choose moduli so that the five-dimensional Newton's
constant is given by $G_5={\pi \over 4}$ and the three tori have equal
volume. When $Q_6=1$ one can compare the singular parts of the
harmonic functions to those of \cite{Bena:2005ni}, and observe that
the integer M2, M5 and KK momentum charges are:
\be
n_I= - { d_I V_R \over 2 }= - { \hat d_I \over 2}\,,\quad N_I={Q_I \over 4}\,,\quad J_{KK} = - { m \over 8 V_R } = - { \hat m \over 8}\,.
\ee
The entropy of the ring is then
\be
S_{BR} = 2 \pi \sqrt{J_4(N_I,n_I,J_{KK})}\,,
\ee
which is exactly the same as for BPS black rings of identical integer 
charges \cite{Bena:2004tk}.

Furthermore, one can use (\ref{dshorizon}) and the limiting values (\ref{limit1})--(\ref{limit4}) to obtain the metric induced on the horizon:
\bea
ds^2_H = \ell^{-4/3} J_4 (d\psi+Q_6 d\phi)^2 + \ell^{2/3}\Bigl[d\theta^2_\Sigma+\sin^2\theta_\Sigma \Bigl(d\phi -{Q_6 \over R^2 V_R^2}(d\psi+Q_6 d\phi)\Bigr)^2\Bigr]\,,
\label{dsnearhorizon}
\eea
where
\be
\ell= {C_{IJK}\over 6} \,\hat{d}_I \hat{d}_J \hat{d}_K \,.
\ee
The factor of ${Q_6 \over R^2 V_R^2}$ in (\ref{dsnearhorizon}) appears
naively to imply that the metric induced on the horizon has conical
singularities at $\theta_{\Sigma}=0$ and $\theta_{\Sigma}= \pi$.
Nevertheless, by carefully investigating the periodicity of $\psi$ and
$\phi$ one can show that the angle that becomes degenerate\footnote{More explicitly, this angle is ${\psi  \over 2 Q_6} - {\phi \over 2} $.} has periodicity $2 \pi$ and hence
no such singularities exist.

\subsection{Asymptotic charges}
\label{chargessec}

To obtain the reduction to four dimensions of the eleven-dimensional  metric (\ref{11D}) one must recast the Gibbons-Hawking $U(1)$ fibration according to:
\bea
ds^2&=& {I_4\over (Z_1 Z_2 Z_3)^{2/3} V^2}\Bigl[d\psi+\vec{A} - {\mu V^2\over I_4} (dt + \vec{\omega})\Bigr]^2 + {V  (Z_1 Z_2 Z_3)^{1/3}\over I_4^{1/2}} ds^2_E \\
&&+ \Bigl({Z_2 Z_3\over Z_1^2}\Bigr)^{1/3} (dx_1^2+dx_2^2)+\Bigl({Z_1 Z_3\over Z_2^2}\Bigr)^{1/3} (dx_3^2+dx_4^2)+\Bigl({Z_1 Z_2\over Z_3^2}\Bigr)^{1/3} (dx_5^2+dx_6^2)\nonumber\,,
\eea
where
\be
ds^2_E = -I_4^{-1/2}(dt+\vec{\omega})^2 + I_4^{1/2} ds^2_3
\ee
is the four-dimensional  Lorentzian metric. In order for this metric to have the canonical normalization at infinity one needs that $I_4\to 1$ at large $r$. This is achieved if one takes
\be
{C_{IJK}\over 6} h\,l_I l_J l_K-m_0^2 =1\,.
\ee
One could also impose that the $\psi$ coordinate be canonically normalized ({\it i.e.} that $g_{\psi\psi}\to 1$ asymptotically) and this requires that
\be
{C_{IJK}\over 6} h^3\, l_I l_J l_K =1\,.
\ee
One can also see that, if $m_0\not=0$, $\mu$ does not vanish at infinity, producing a non-vanishing $g_{t\psi}$. This means that one is in a rotating frame at infinity, which can be undone by a re-definition of the coordinate $\psi$, as
\be
\tilde \psi = \psi + h m_0 \,t\,.
\ee
In terms of $\tilde \psi$ the metric is explicitly asymptotically flat and it is straightforward to compute the associated asymptotic charges. The M2 charges are:
\be
\hat{Q}_I = Q_I + {Q_6\over R^2}  {C_{IJK}\over 2} d_J d_K \,,
\ee
while the KK-monopole charge is simply given by $Q_6$ and the M5 charges by $d_I$. The mass is given by the BPS-like formula:
\be
M= {C_{IJK}\over 6}{l_I l_J l_K\over 4} Q_6 + {h\over 4} {C_{IJK}\over 2} \hat{Q_I} l_J l_K-{m_0 \,h\over 2} l_I d_I\,.
\ee
Note that here $Q_6$ and $\hat{Q}_I$ denote the absolute values of the charges.

The momentum along the KK direction $\tilde\psi$ is:
\be
P= h^2 \Bigl({C_{IJK}\over 6}h\,l_I l_J l_K +m_0^2\Bigr)\,l_I d_I -m_0\, h^2 \,{C_{IJK}\over 2} \hat{Q_I} l_J l_K -m_0^3\, Q_6\,,
\ee
and the angular momentum in the non-compact $\mathbb{R}^3$ is:
\bea
J &=& R \Bigl(m- h{l_I d_I\over 2}\Bigr)+{Q_6\over 2 R} d_I Q_I + {Q_6^2\over R^3} {C_{IJK}\over 6} d_I d_J d_K\nn
&=& {Q_6\over 2} l_I d_I   +{Q_6\over 2 R} d_I Q_I + {Q_6\over 2 R^2}\Bigl(h+{2 Q_6\over R}\Bigr){C_{IJK}\over 6}  d_I d_J d_K\,.
\eea

If $m_0=0$ and the $l_I$ and $h$ are equal to 1, the mass formula
takes a more familiar form, as a sum of absolute values of charges:
\be
M= {Q_6\over 4} + {1 \over 4} \sum_I \hat{Q_I}\,,
\ee
and the KK momentum along the GH fiber is just the sum of the dipole
charges (much like for BPS black rings):
\be P= \sum_I d_I = \sum_I {\hat d_I
  \over 1+ Q_6/R}\,.
\ee

Moreover, the four-dimensional angular momentum becomes
\be
J = {Q_6 P \over 2}
+{Q_6\over 2 R} d_I Q_I + {Q_6\over 2 R^2}\Bigl(1+{2 Q_6\over 
  R}\Bigr) {C_{IJK}\over 6}  d_I d_J d_K\,,
\ee
where now we can identify the first piece as coming from the Poynting
vector caused by the KK electric and magnetic charges and the other
pieces as coming from the interactions between the electric M2 charges
and the magnetic M5 charges.  When the black ring becomes a supertube
($d_1=d_2=Q_3=0$), the latter interactions are zero, and the KK
Poynting term $ {Q_6 P \over 2}$ is the only one that survives.

\section{Almost BPS supertubes}

\subsection{The supertube solution}

From a supergravity perspective, a supertube \cite{supertube} can be
thought of as a particular black ring with only two charges and one
dipole charge. One can thus trivially obtain an ``almost BPS''
supertube from the non-BPS solution above taking the following
harmonic functions
\bea
&&K_1 = K_2 = 0\,,\quad K_3 = {d_3\over \Sigma}\, \quad V= 1+ {Q_6 \over r} \\
&&L_1 = 1+{Q_1\over \Sigma}\,,\quad L_2=1+{Q_2\over \Sigma}\,,\quad L_3=1\,,\\
&&M = m_0 + {m\over \Sigma}+{\tilde m\over r}\,.
\eea
The solution simplifies considerably, and one finds
\bea
&&a_1 = a_2 =0\,,\quad a_3 = K_3 (d\psi+\vec{A})+ \vec{a}_3\,,\quad *_3 d\vec{a}_3 = V dK_3 - K_3 dV\nn
 &&\Rightarrow\,\,\vec{a}_3 = d_3 {r\cos\theta-R\over \Sigma} d\phi + Q_6 d_3 {r-R\cos\theta\over R\Sigma} d\phi\,,\nn
&&Z_I = L_I\label{almostbpstube}\\
&&\mu = {M\over V}+{1\over 2} K_3\,,\quad *_3 d \vec{\omega} = -dM + {1\over 2} (V dK_3 - K_3 dV)\nn
&&\Rightarrow\,\, \vec{\omega} = \Bigl(- m+{d_3\over 2}\Bigr) {r \cos\theta - R\over \Sigma}\, d\phi -{\tilde m} \cos\theta\,d\phi + {Q_6 d_3\over 2} {r-R \cos\theta\over R\, \Sigma} d\phi\,.\nonumber
\eea

The supertube is smooth in a duality frame in which the electric (M2) charges correspond to D1 and D5 branes and the magnetic (M5) dipole moment corresponds to a KK-monopole wrapped around the Taub-NUT direction. In this frame, the ten-dimensional  string metric is:
\be
ds^2 = -{1\over \sqrt{Z_1 Z_2} Z_3}(dt+k)^2 + {Z_3\over \sqrt{Z_1 Z_2}} \Bigl(dy+a_3 - {dt+k\over Z_3}\Bigr)^2 +
\sqrt{Z_1 Z_2}ds^2_4 + \sqrt{Z_2\over Z_1} \sum_{a=1}^4 dx_a^2\,,
\label{D1D5}
\ee
where $y$ the common  D1-D5 direction.  Standard BPS supertubes are regular in this frame and so we now consider the regularity of the metric of the ``almost BPS''  supertubes.  The coefficient of $d\psi^2$ in the metric is:
\bea
g_{\psi\psi}&=&{1\over \sqrt{Z_1 Z_2}}(Z_3 a_{3,\psi}^2 - 2 \mu a_{3,\psi}+Z_1 Z_2 V^{-1})\nn
&=&  {1\over V \sqrt{L_1 L_2}} (L_1 L_2 - 2 M K_3)\,,
\eea
where in the second line we have used the expressions for $Z_I$, $a_3$ and $\mu$ given in (\ref{almostbpstube}). The requirement that $g_{\psi\psi}$ be finite for $\Sigma\to 0$ implies
\be
m={Q_1 Q_2\over 2 d_3}\,.
\ee
In order for $\vec{\omega}$ not to have any Dirac-Misner string pathologies around the point $\Sigma=0$ it is necessary that $\vec{\omega}$ vanish for $\theta=0$ and $r$ greater or smaller than $R$. These conditions imply:
\be
m= {V_R d_3\over 2}\quad \mathrm{with} \quad V_R = 1+{Q_6\over R}\,.
\ee
Combining these two relations for $m$ one obtains an equation that determines the supertube location $R$:
\be
V_R = {Q_1 Q_2\over d_3^2}\,.
\ee
Finally one should look at regularity at the Taub-NUT center $r=0$. As the coordinate $\psi$ degenerates at $r=0$, $\mu$ must vanish to prevent CTC's, which implies
\be
{\tilde m}=- {d_3 Q_6\over 2 R}\,.
\ee

\subsection{Comparing BPS and ``almost BPS'' supertubes}

Having found a smooth supertube metric that solves the ``almost BPS''
equations (\ref{thetaeq})--(\ref{keq}), we can compare it to that of a
BPS supertube, and show that despite their rather different
appearance, the two solutions are identical.

Denoting with a ``hat'' the quantities associated with the BPS
solution, we recall that the BPS supertube solution is given by:
\bea
&&\hat{a}_3 = {\hat{K}_3\over
  V}(d\psi+\vec{A})+\hat{\vec{a}}_3\,,\quad *_3 d \hat{\vec{a}}_3 = -
d\hat{K}_3\,,\quad \hat{Z}_I = \hat{L}_I\,, \nn
&&\hat{k}= \Bigl(\hat{M} +
{\hat{K}_3\over 2V}\Bigr)(d\psi+\vec{A}
d\phi)+\hat{\vec{\omega}}\,,\quad *_3 d\hat{\vec{\omega}} = V d{\hat
  M} - \hat{M} dV -{1\over 2} d{\hat K}_3\,.
\label{bpstube}
\eea

Since the supertube solution has $Z_3=1$, one can absorb the term
${\displaystyle - {dt / Z_3}}$ in equation (\ref{D1D5}) by the
coordinate shift $y\to y + t$. Thus the dipole potential $a_3$ only
enters in the metric via the combination $(dy + a_3-k)^2$.
Comparing the BPS
expressions (\ref{bpstube}) to the ``almost BPS'' ones
(\ref{almostbpstube}), one sees that, under the identifications
\be
{\hat K}_3 = 2 M \,,\quad {\hat M}={K_3\over 2}\,,\quad {\hat L}_I = L_I
\ee
one has
\be
\hat{a}_3 - \hat{k}= -(a_3 - k)\,,\quad \hat{Z}_I= Z_I\,,\quad  \hat k = k\,.
\ee
Hence, the BPS and ``almost BPS'' supertube solutions can be related to each other by  flipping the sign of $y$
and interchanging harmonic functions.

\section{General extremal non-BPS rotating black holes}
\label{bh0sec}

In this section we present the other main result of this paper: a
rotating five-charge extremal non-BPS black hole in four dimensions.
This black hole can serve as the seed solution for the most generic
under-rotating non-BPS extremal black hole in the $STU$ model and in
${\cal N}=8$ supergravity in four dimensions, and can be thought of as
coming from the non-BPS extension of the five-dimensional BPS rotating
(BMPV) black hole to an asymptotically Taub-NUT solution\footnote{For
  a recent discussion of the BPS extension of this black hole to
  Taub-NUT see \cite{4D5D}.}.

We first construct and analyze this black hole, and then show that for
special values of the charges it reduces to the under-rotating D0-D6
extremal black hole \cite{rasheed-larsen}.

\subsection{The solution}

The harmonic functions associated with the KK-monopole and electric (M2) charges have the usual form
\be
V = h+{Q_6\over r}\,,\quad L_I = 1+{Q_I\over r}\,,
\label{vlbh}
\ee
where for simplicity we have set to one the constants $l_I$ in the $L_I$ 
harmonic functions. The solution with arbitrary moduli is presented in Section \ref{bhsec}.

The dipole charges vanish, and hence $K_I=0$. The harmonic function, $M$, which encodes the angular momentum of the solution is taken to have the form:
\be
M= m_0 + {m\over r}+\alpha {\cos\theta\over r^2}\,.
\ee
The term proportional to $\alpha$ is the harmonic potential is sourced by a dipole  at the origin of Taub-NUT space and, as we will see, is needed to generate the angular momentum of the black hole.

With this choice of harmonic functions, the ``almost BPS'' equations (\ref{thetaeq}-\ref{keq}) are solved
by\footnote{The vector potential $\vec{\omega}$ dual to the dipole field ${\cos\theta\over r^2}$ follows from the identity
\be
*_3 d\Bigl( {\sin^2\theta\over r} d\phi\Bigr) = - d \Bigl({\cos\theta\over r^2}\Bigr)\,. \nonumber
\ee}
\bea
\Theta_I =0\,,\quad Z_I = L_I\,,\quad \mu = {M\over V}= {m_0\over V} + {m\over V r}+\alpha {\cos\theta\over V r^2}\,,\quad \vec{\omega} = - m \cos\theta d\phi +\alpha {\sin^2\theta\over r} d\phi\,.
\eea

Absence of Dirac-Misner strings requires that $\vec{\omega}$ vanish both at $\theta=0$ and $\theta=\pi$, and hence we must take
\be
m=0\,.
\ee
Nevertheless, $\alpha$ remains as a free parameter of the solution and it encodes the angular momentum.
To see this more explicitly we compute the conserved charges. As shown in  section \ref{chargessec}, the four-dimensional 
Lorentzian metric is:
\be
ds^2_E = -I_4^{-1/2}(dt+\vec{\omega})^2+I_4^{1/2} ds^2_3\,,\quad I_4 = Z_1 Z_2 Z_3 V - \mu^2 V^2\,, \label{BHSOLUTION}
\ee
and the electric component of the KK gauge field coming from the reduction along the Taub-NUT fiber is
\be
A_{KK}= -{\mu V^2\over I_4}\,.
\ee
The normalization condition $I_4\to 1$ for large $r$ requires
\be
h-m_0^2=1\,.
\ee
The KK momentum along $\psi$, found from the asymptotic expansion of $A_{KK}$, is
\be
P = m_0 (h^2 (Q_1 + Q_2 + Q_3)+ m_0^2 Q_6)\,,
\ee
and the $\mathbb{R}^3$ angular momentum, encoded in $\vec{\omega}$,  is
\be
J = \alpha\,.
\ee

One can also show that this solution has a regular horizon of finite area. In the near-horizon ($r \to 0$) limit, one has
\be
I_4 \to {Q_1 Q_2 Q_3 Q_6-\alpha^2 \cos^2\theta\over r^4}\,,\qquad
\omega_\phi \to  \alpha {\sin^2\theta\over r}\,,
\ee
and thus the volume element of the metric induced on the horizon is
\be
\sqrt{g_H}= r (I_4 r^2 \sin^2\theta - \omega^2_\phi)^{1/2} \approx \sin\theta (Q_1 Q_2 Q_3 Q_6-\alpha^2)^{1/2}\,.
\ee
The horizon area is
\be
A_H = (4\pi Q_6)(4\pi)\sqrt{Q_1 Q_2 Q_3 Q_6-\alpha^2}\,,
\ee
which coincides with the area of the corresponding BMPV black hole.

\subsection{The extremal rotating D0-D6 black hole}
\label{RL}

We now discuss the relationship between the solution presented
above to the one of Rasheed and Larsen \cite{rasheed-larsen}.  First of
all, the solution of Rasheed and Larsen can be compared to ours only
in the ``slowly rotating'' or ``ergo-free'' extremal limit: $a \to 0$,
$m\to 0$, keeping $a/m = J $ fixed.  In this limit the metric of
\cite{rasheed-larsen}  can be recast in a form similar to the one of
(\ref{BHSOLUTION}):
\begin{equation}
	ds^2 = - \frac{r^2}{\sqrt{H_1 H_2}}\left(dt + {\mathbf B}\right)^2 + \frac{\sqrt{H_1 H_2}}{r^2} ds_3^2, \label{Larsen1}
\end{equation}
where
\begin{equation}
	{\mathbf B} = \frac{(pq)^{3/2}}{2(p+q)} \, J \, \frac{\sin^2 \theta}{r} \, d \phi,
\end{equation}
\begin{equation}
	H_1 = r^2 + r p + \frac{p^2 q}{2(p + q)} - \frac{p^2 q}{2(p + q)} J \cos \theta,
\end{equation}
\begin{equation}
	H_2 = r^2 + r q + \frac{q^2 p}{2(p + q)} + \frac{q^2 p}{2(p + q)} J \cos \theta.
\end{equation}
This solution has a single scalar field running
\begin{equation}
	z = i \sqrt{\frac{H_2}{H_1}} \label{singlescalar}
\end{equation}
and a vanishing axion.
The physical D0 and D6 charges $Q$ and $P$ are related to $p$ and $q$ by
\begin{equation}
	Q^2 = \frac{q^3}{4(p+q)}, \quad	P^2 = \frac{p^3}{4(p+q)}.
\end{equation}

This solution is related, by a U-duality transformation,  to the solution presented
above.  We will establish this by applying an appropriate transformation to the scalar field (\ref{singlescalar})
and showing that the resulting fields and charges fall in a special subset
of those presented above.  Since we are starting from a special
configuration with only two charges turned on and no axion, we do not
expect to be able to generate the most general solution, but we will
obviously obtain some constraints on the allowed values for the moduli
at infinity.

In order to simplify computations, we consider the ${\cal N}=2$
truncation of the M-theory description used earlier.  Hence we will look
at compactifications on $T^6/(Z_2 \times Z_2) \times S^1$, where the
last $S^1$ is parametrized by $\psi$ and the orbifold action is the
trivial one preserving the 2-forms  $dx_1 \wedge dx_2$, $dx_3
\wedge dx_4$ and $dx_5 \wedge dx_6$.  The resulting ${\cal N}=2$ effective
theory is described by an $STU$ model, with scalar fields in the
vector multiplets parametrizing:
\begin{equation}
	\left[\frac{SU(1,1)}{U(1)}\right]^3 \simeq \frac{SU(1,1)}{U(1)} \times \frac{SO(2,2)}{SO(2) \times SO(2)}.
\end{equation}
The three complex moduli for our solution are given by
\begin{equation}
	t_I = {4M \over V Z_I} + 4 i \, {\rm e}^{-\phi} B_I,
\label{moduli-met}
\end{equation}
where
\begin{equation}
	B_I = \frac{(\frac12 C_{IJK} Z_J Z_K)^{1/3}}{Z_I^{2/3}},
\end{equation}
and the dilaton is
\begin{equation}
	{\rm e}^{-2 \phi} = \frac{I_4}{(Z_1 Z_2 Z_3)^{2/3} V^2}.
\end{equation}
The duality action on the three scalar fields then acts as follows:
\begin{equation}
	t_I \to  \frac{a_I t_I + b_I}{c_I t_I + d_I} \qquad {\rm (no \  sum)}
\end{equation}
where
\begin{equation}
	M_I = \left(\begin{array}{cc}
	  a_I & b_I\\
	  c_I & d_I
	\end{array}\right),
\end{equation}
are SL($2,{\mathbb R}$) matrices.

Without rotation one can immediately check that our solution reduces to
\begin{equation}
	t_I = \frac{4}{V Z_I}(m_0 + i \, {\rm e}^{-2 U}),
\end{equation}
with ${\rm e}^{-2 U} = \sqrt{I_4}$, which is the one presented in Equation (4.34) of \cite{LopesCardoso:2007ky}.
This is easily dualized to the generating solution by \cite{Gimon:2007mh} by taking
\begin{equation}
	M_I =  \left(\begin{array}{cc}
	 0 & 1\\
	 -1 & 0
	\end{array}\right)
\end{equation}
which yields
\begin{equation}
	t_I = \frac{1}{2 \,C_{IJK}Z_J Z_K}(m_0-i\, {\rm e}^{-2 U}).
\end{equation}
At this point one can further dualize to D0-D6 charges by
following the duality rotations described in \cite{Gimon:2007mh}.  The complete duality transformation mapping the
D6-D2-D2-D2 system into the D0-D6 is then given by
\begin{equation}
	M_I = - \frac{1}{\sqrt{2 \lambda \rho_I}}\left(\begin{array}{cc}
		 -\rho_I & 1\\
	  - \rho_I \lambda & - \lambda
	\end{array}\right), \label{Mi}
\end{equation}
where
\begin{equation}
	\lambda = \left(\frac{P}{Q}\right)^{1/3}, 
\quad \rho_I = \sqrt{\frac{p^0 q_I}{\frac12 C_{IJK} q_J q_K}}, \label{condlambda}
\end{equation}
with $16 p^0 = Q_6$, $q_I = Q_I$ and $(P Q)^2 = 4 p^0 q_1 q_2 q_3$.

Following the inverse route, we can start from (\ref{Larsen1})--(\ref{singlescalar}) and apply the inverse transformation:
\begin{equation}
	M_I = - \frac{1}{\sqrt{2 \lambda \rho_I}}\left(\begin{array}{cc}
		 -\lambda & -1\\
	   \rho_I \lambda & - \rho_I
	\end{array}\right). \label{Minv}
\end{equation}
The four-dimensional dilaton can be identified to the diagonal scalar $t_1 = t_2 = t_3 = z$.
After applying the duality transformation we obtain
\begin{equation}
	t_I = -\frac{1}{\rho_I} \frac{\lambda z + 1}{\lambda z - 1}
\end{equation}
which we expect to match the moduli of our metric (\ref{moduli-met}),
which become\footnote{As in \cite{Gimon:2007mh}, we use conventions in which 
$|t_I| = {1 \over \rho_I}$ at infinity.}
\begin{equation}
	t_I = \frac{4}{V Z_I} \left(\mu V + i \, \sqrt{I_4}\right).
\end{equation}
Using the explicit expression for $z$ given in (\ref{singlescalar}) we can see that one needs to identify
\begin{equation}
	V Z_I = \frac{2 \rho_I}{\lambda} \frac{H_1 + \lambda^2 H_2}{r^2}
\end{equation}
and
\begin{equation}
	V \mu = \frac{1}{2 \lambda} \frac{H_1 - \lambda^2 H_2}{r^2}.
\end{equation}
This can be achieved for
\begin{equation}
	\lambda = \sqrt{\frac{p}{q}}, \quad \rho_I = \frac{p+q}{2 (p q)^{3/2}} \, Q_6 q_I,
\end{equation}
which is equivalent to (\ref{condlambda}) and
\begin{equation}
	h = \frac{p + q}{p q} Q_6, \quad l_I = \frac{p + q}{p q}q_I, \quad m_0 = \frac{q-p}{2 \sqrt{pq}}, \quad \alpha = - \frac{(pq)^{3/2}}{2 (p+q)} J,
\end{equation}
where the $l_I$ are the constants in the harmonic functions $L_I$,
which, for simplicity, we have set to one in equation (\ref{vlbh}), but
which we will explicitly include in the next section (see equation
(\ref{lIbh})) when discussing the general black-hole-black-ring
solution.

Hence for special values of the charges and of the moduli, our solution
can be dualized to the under-rotating extremal limit of the D0-D6
Rasheed-Larsen black hole. However, our solution has generic charges and 
moduli and hence it is more general; its duality orbit includes all the
under-rotating extremal black hole solutions of the $STU$ model or of
${\cal N}=8$ supergravity in four dimensions.

\section{Non-BPS  black ring in a black-hole background}
\label{bhsec}
%
Making use of the linear structure underlying the equations
(\ref{thetaeq})--(\ref{keq}), it is possible to superimpose the solutions
constructed in the previous sections to generate the metric describing
a non-BPS black ring with a rotating black hole at the origin of
Taub-NUT space. Starting from the black ring solution of section
\ref{ringsec}, adding the rotating black hole corresponds to adding a
$1/r$ term to the harmonic functions $L_I$, which therefore becomes
\be
L_I = l_I+{Q_I\over \Sigma} + {\tilde Q_I\over r}\,,
\label{lIbh}
\ee
and a ``dipole'' source centered at $r=0$ to the harmonic function $M$:
\be
M = m_0 + {m\over \Sigma}+{\tilde m\over r} + {\tilde \alpha}{\cos \theta_{\Sigma} \over \Sigma^2} +\beta {\cos\theta\over r^2}\,.
\ee
The dipole potentials $a_I$ are left untouched, and  are still given by the expressions in (\ref{dipolenonbpsring}). The warp factors $Z_I$ are obtained by replacing the old  functions $L_I$ with the new ones given in (\ref{lIbh}):
\be
Z_I = l_I+{Q_I\over \Sigma}+ {\tilde Q_I\over r}+{C_{IJK}\over 2} {d_J d_K\over \Sigma^2} \Bigl(h+ {Q_6 r\over R^2}\Bigr)\,.
\ee
The new $1/r$ term in $Z_I$ adds the contribution
\be
\Bigl(h+{\tilde Q_6\over r}\Bigr){\tilde Q_I\over r}d\Bigl({d_I\over \Sigma}\Bigr)
\ee
to the r.h.s. of the equation for $k$ (\ref{keqring}). Hence $k$ receives two new contributions. The first one is given by the solution of
\be
d(V \,\mu_7) + *_3 d\vec{\omega}_7 = h {\tilde Q_I\over r} d\Bigl({d_I\over \Sigma}\Bigr)\,.
\ee
This equation is easily solved by
\be
\mu_7 = {h\,{\tilde Q}_I d_I\over 2 V\,r \Sigma} \,,\quad \vec{\omega}_7 = {h \,{\tilde Q}_I d_I\over 2}{r-R\cos\theta\over R \Sigma} d\phi\,.
\ee
The other new term in $k$ is found by solving
\be
d(V \,\mu_8) + *_3 d\vec{\omega}_8 =  {Q_6 \,\tilde Q_I\over r^2} d\Bigl({d_I\over \Sigma}\Bigr)\,.
\ee
Again one can find the solution by using the corresponding solution for a flat base. The result is
\be
\mu_8 = {Q_6 \,\tilde Q_I d_I\over R\,V\,r\,\Sigma}\,\cos\theta\,,\quad \vec{\omega}_8 = {Q_6 \,\tilde Q_I d_I\over R\,\Sigma}\,\sin^2\theta\, d\phi\,.
\ee
Furthermore the term proportional to $\beta$ in $M$ generates an extra contribution given by
\be
\mu_9 = \beta {\cos\theta\over V r^2}\,,\quad \vec{\omega}_9 = \beta {\sin^2\theta\over r} d\phi\,.
\ee
Adding the new terms to the previous black ring result, one finds the full solution for $k$:
\bea
\mu &\!\!\!=\!\!\!&  {m_0\over V}+ {m\over V \,\Sigma}+ {\tilde m\over V\,r }+\beta {\cos\theta\over V r^2}+{l_I d_I\over 2 \Sigma}+{h Q_I d_I\over 2V\, \Sigma^2}+Q_6 Q_I d_I {\cos\theta\over 2 R \,V \Sigma^2}+ {h\,{\tilde Q}_I d_I\over 2 V\,r \Sigma}+ {Q_6 \,\tilde Q_I d_I\over R\,V\,r\,\Sigma}
\\&&+{C_{IJK}\over 6} d_I d_J d_K\Bigl[\Bigl(h^2+{Q_6^2\over R^2} \Bigr)  \Bigl( {r\cos\theta\over R\,V\, \Sigma^3}+\alpha {r\cos\theta-R\over R\,V\, \Sigma^3}\Bigr)+Q_6  h  {3 r^2 + R^2\over 2 R^2 V\,r \,\Sigma^3}\Bigr]\,,\nn
\vec{\omega}&\!\!\!=\!\!\!& \Bigl\{\kappa- m {r \cos\theta- R\over \Sigma}-{\tilde m} \cos\theta +\beta {\sin^2\theta\over r} d\phi+ {h l_I d_I\over 2}{r\cos\theta-R\over \Sigma} +{Q_6 l_I d_I\over 2}{r-R\cos\theta\over R \Sigma} \nn&&
+ Q_6 Q_I d_I {r\sin^2\theta\over 2 R \,\Sigma^2} + {h \,{\tilde Q}_I d_I\over 2}{r-R\cos\theta\over R \Sigma}+ {Q_6 \,\tilde Q_I d_I\over R\,\Sigma}\,\sin^2\theta\nn
&&+{C_{IJK}\over 6} d_I d_J d_K\Bigl[\Bigl(h^2+{Q_6^2\over R^2} \Bigr) (1+\alpha) {r^2 \sin^2\theta\over R\, \Sigma^3}+ Q_6 h  {r (3 R^2 + r^2)- R (3 r^2 + R^2) \cos\theta\over 2 R^3\, \Sigma^3}\Bigr]\Bigr\} d\phi\,.\nonumber
\eea

The absence of Dirac-Misner strings requires that $\vec{\omega}$ vanishes on the $z$ axis. This imposes the following constraints, which are the generalization of (\ref{regularityring})
\bea
m &=& \Bigl(h+{Q_6\over R}\Bigr){l_I d_I\over 2}+{C_{IJK}\over 6} {Q_6 h  d_I d_J d_K\over 2 R^3}+ {h\over 2R} {\tilde Q}_I d_I\nn
{\tilde m}&=& \kappa = -Q_6 \Bigl({l_I d_I\over 2 R}+ {C_{IJK}\over 6} {h d_I d_J d_K\over 2 R^3}\Bigr)-{h\over 2R} {\tilde Q}_I d_I\,.
\eea
The first equation can again be thought of as the generalization of
the bubble equations
\cite{denef,Bena:2005va,Berglund:2005vb,Saxena:2005uk} to the most
generic two-center non-BPS extremal solution.

The topology of the black ring horizon at $\Sigma=0$ is not
affected by the black hole. As above, if $\alpha$ is chosen as in (\ref{alphavalue}), this solution has horizon of finite area at $\Sigma=0$ with an $S^2\times S^1$ geometry.  The area of this horizon is:
\be
A_H = 16 \pi^2 Q_6\,{\tilde J}_4^{1/2} \,,
\ee
where
\be
{\tilde J}_4^{1/2} = {1\over 2}\sum_{I<J} \hat{d}_I\hat{d}_J\,Q_I  Q_J -{1\over 4}\sum_I  \hat{d}_I^2\, Q_I^2 -{C_{IJK}\over 6} \hat{d}_I \hat{d}_J \hat{d}_K\Bigl(2 \hat{m} \,+ {Q_6\over R^2\,V_R^2}  {\tilde Q}_I \hat{d}_I\Bigr)\,.
\label{mcomb}
\ee

As for BPS black rings in black-hole backgrounds \cite{Bena:2005zy},
the integer D0 charge of the ring is no longer proportional to $\hat
m$ but rather to the combination that appears in equation
(\ref{mcomb}): 
\be 
\hat{m} \,+ {Q_6\over 2 R^2\,V_R^2} {\tilde Q}_I
\hat{d}_I\,.  
\ee

The black hole at the center of the Taub-NUT space has five-dimensional horizon
area equal to:
\be
A_{BH}=(4\pi Q_6)(4\pi)\sqrt{Q_6 {C_{IJK}\over 6} \tilde Q_I \tilde
Q_J \tilde Q_K-\beta^2}\,.
\ee
This black hole carries electric D6 and D2 charges ($Q_6$ and $\tilde Q_I$),
and angular momentum $\beta$.

\section{Conclusions and future directions}

We have explicitly constructed three-charge three-dipole charge
extremal non-BPS black rings in Taub-NUT, both in the absence and in
the presence of a three-charge black hole. These rings are locally
identical to the supersymmetric black rings, but break supersymmetry
because the D6 brane that can be thought of as sourcing the Taub-NUT
space has a reversed orientation compared to the BPS embedding.
Our solutions become identical to the BPS rings both in the limit when
Taub-NUT becomes $\IR^4$ and in the limit when it becomes $\IR^3
\times S^1$, where the orientation of the D6 brane becomes irrelevant.

We have also constructed the solution for the non-BPS embedding of a
two-charge supertube, and have shown that this solution is the same as
that of a BPS supertube, despite the rather different form of the
ingredients that enter in its construction. This agrees with the
intuition that supersymmetry is broken by the incompatible
supersymmetry constraints imposed by multiple branes. When only two D2
charges are present flipping the charge of the D6 brane creates no
such incompatibility: There are still consistent supersymmetries with
all three branes.

We have also found an extremal rotating non-BPS five-charge
(D6-D2-D2-D2-D0) black hole in four dimensions. This solution is the
seed for the most general extremal (under)rotating non-BPS black hole
solution of the $STU$ model or of ${\cal N}=8$ supergravity in four
dimensions. For particular values of the charges and moduli we have
shown that this solution can be dualized to the Kaluza Klein rotating
black hole solution of Rasheed and Larsen \cite{rasheed-larsen} or its
U-duals \cite{Giusto:2007tt}.

Using our solution-generating method we have also constructed a
solution that contains both a rotating black hole and a black ring.
This solution descends in four dimensions to a two-black-hole non-BPS
bound state, where one of the black holes has five charges
(D6-D2-D2-D2-D0) and the other has seven charges
(D4-D4-D4-D2-D2-D2-D0) \footnote{Note that that the charges at the two
  centers are mutually nonlocal, and hence this solution is more
  general than the one constructed in \cite{Gaiotto:2007ag}.}. The
bubble equations that determine the distance between the two black
holes are cubic in this distance, and hence are more complicated than
those governing BPS multi-center
solutions \cite{denef,Bena:2005va,Berglund:2005vb,Saxena:2005uk}.

This two-center solution appears to be the most general one can
construct within the framework of \cite{Goldstein}. Furthermore, the
charges of its centers can be dualized to those of more generic
extremal non-BPS two-black-hole configurations. It would be
interesting to further investigate how generic these charges are, and
whether our solution lies in the duality orbit of the most generic
two-center extremal solution.

As we discussed in the Introduction, our work bridges the gap between
two rather disconnected bodies of research -- the construction of
extremal supergravity solutions using fake superpotentials
\cite{Ceresole:2007wx,LopesCardoso:2007ky,Andrianopoli:2007gt,Gimon:2007mh,Bellucci:2008sv}
and the embedding of five-dimensional black holes and black rings in
Taub-NUT \cite{Ford:2007th,Giusto:2007fx,Camps:2008hb}. It would be
interesting to see whether the fake superpotential approach can be
extended to describe multi-center solutions, and whether the
two-center solution we obtain is the most general one can find within
this framework.  It would be equally interesting to see if the
construction in \cite{Camps:2008hb} can be extended to electrically
and magnetically-charged black rings in Taub-NUT, and whether the
extremal limit of these rings can be compared to the extremal non-BPS
rings we construct.

Another very important extension of this work would be to encompass
families of multi-center solutions. The method we have outlined here
allows one to recycle a considerable part of the known BPS
multi-center solutions in $\IR^4$ and in $\IR^3 \times S^1$. Indeed,
just by writing these solutions as non-BPS solutions one can read off
the warp factors, as well as all the angular momentum terms that
are not cubic in the dipole charges. Nevertheless, the terms cubic in
the dipole charges (that satisfy an equation similar to (\ref{mu5eq}))
appear to be somewhat harder to obtain.

Last, but not least, it would be interesting to use the ansatz of
\cite{Goldstein} to extend the construction of BPS smooth multi-center
bubbling solutions \cite{Bena:2005va,Berglund:2005vb,Saxena:2005uk} to
non-BPS smooth extremal solutions, which would correspond to
microstates of extremal non-BPS black holes.  It would be particularly
interesting if non-BPS solutions exhibited scaling behavior.

\bigskip
\leftline{\bf Acknowledgments}
\smallskip
We would like to thank R. D'Auria and M. Trigiante for interesting
discussions.  The work of IB, CR and SG was supported in part by the
DSM CEA-Saclay, by the ANR grants BLAN 06-3-137168 and JCJC ERCS07-12,
and by the Marie Curie IRG 046430. The work of GD is partially
supported by an excellence grant funded by ``Fondazione CARIPARO''.
The work of NPW was supported in part by DOE grant DE-FG03-84ER-40168.




\begin{thebibliography}{99}



\bibitem{Bena:2004de}
  I.~Bena and N.P.~Warner,
``One ring to rule them all ... and in the darkness bind them?,''
  Adv.\ Theor.\ Math.\ Phys.\  {\bf 9}, 667 (2005)
  [arXiv:hep-th/0408106].

\bibitem{Gutowski:2004yv}
  J.~B.~Gutowski and H.~S.~Reall,
  JHEP {\bf 0404}, 048 (2004)
  [arXiv:hep-th/0401129].



\bibitem{Gauntlett:2002nw}
  J.~P.~Gauntlett, J.~B.~Gutowski, C.~M.~Hull, S.~Pakis and H.~S.~Reall,
``All supersymmetric solutions of minimal supergravity in five dimensions,''
  Class.\ Quant.\ Grav.\  {\bf 20}, 4587 (2003)
  [arXiv:hep-th/0209114].


\bibitem{Gauntlett:2004qy}
  J.~P.~Gauntlett and J.~B.~Gutowski,
``General concentric black rings,''
  Phys.\ Rev.\ D {\bf 71}, 045002 (2005)
  [arXiv:hep-th/0408122].


\bibitem{Bena:2005ni}
  I.~Bena, P.~Kraus and N.~P.~Warner,
  ``Black rings in Taub-NUT,''
  Phys.\ Rev.\  D {\bf 72}, 084019 (2005)
  [arXiv:hep-th/0504142].

\bibitem{denef}
 F.~Denef,
  ``Supergravity flows and D-brane stability,''
  JHEP {\bf 0008}, 050 (2000)
  [arXiv:hep-th/0005049].

  F.~Denef,
  ``Quantum quivers and Hall/hole halos,''
  JHEP {\bf 0210}, 023 (2002)
  [arXiv:hep-th/0206072].

  B.~Bates and F.~Denef,
  ``Exact solutions for supersymmetric stationary black hole composites,''
  [arXiv:hep-th/0304094].

\bibitem{Goldstein}
  K.~Goldstein and S.~Katmadas,
  ``Almost BPS black holes,''
  arXiv:0812.4183 [hep-th].


\bibitem{BenaKrausKKM}
  I.~Bena and P.~Kraus,
  ``Microstates of the D1-D5-KK system,''
  Phys.\ Rev.\  D {\bf 72}, 025007 (2005)
  [arXiv:hep-th/0503053].

\bibitem{Bena:2004wv}
  I.~Bena,
``Splitting hairs of the three charge black hole,''
  Phys.\ Rev.\ D {\bf 70}, 105018 (2004)
  [arXiv:hep-th/0404073].

\bibitem{Elvang:2004rt}
  H.~Elvang, R.~Emparan, D.~Mateos and H.~S.~Reall,
``A supersymmetric black ring,''
  Phys.\ Rev.\ Lett.\  {\bf 93}, 211302 (2004)
  [arXiv:hep-th/0407065].


\bibitem{Elvang:2004ds}
  H.~Elvang, R.~Emparan, D.~Mateos and H.~S.~Reall,
  ``Supersymmetric black rings and three-charge supertubes,''
  Phys.\ Rev.\  D {\bf 71}, 024033 (2005)
  [arXiv:hep-th/0408120].

\bibitem{Elvang:2005sa}
  H.~Elvang, R.~Emparan, D.~Mateos and H.~S.~Reall,
  ``Supersymmetric 4D rotating black holes from 5D black rings,''
  JHEP {\bf 0508}, 042 (2005)
  [arXiv:hep-th/0504125].

\bibitem{Gaiotto:2005xt}
  D.~Gaiotto, A.~Strominger and X.~Yin,
  ``5D black rings and 4D black holes,''
  JHEP {\bf 0602} (2006) 023
  [arXiv:hep-th/0504126].

\bibitem{Bena:2004tk}
  I.~Bena and P.~Kraus,
  ``Microscopic description of black rings in AdS/CFT,''
  JHEP {\bf 0412}, 070 (2004)
  [arXiv:hep-th/0408186].

\bibitem{LopesCardoso:2007ky}
  G.~Lopes Cardoso, A.~Ceresole, G.~Dall'Agata, J.~M.~Oberreuter and J.~Perz,
  ``First-order flow equations for extremal black holes in very special
  geometry,''
  JHEP {\bf 0710}, 063 (2007)
  [arXiv:0706.3373 [hep-th]].


\bibitem{Gimon:2007mh}
  E.~G.~Gimon, F.~Larsen and J.~Simon,
  ``Black Holes in Supergravity: the non-BPS Branch,''
  JHEP {\bf 0801}, 040 (2008)
  [arXiv:0710.4967 [hep-th]].

  K.~Hotta and T.~Kubota,
  ``Exact Solutions and the Attractor Mechanism in Non-BPS Black Holes,''
  Prog.\ Theor.\ Phys.\  {\bf 118}, 969 (2007)
  [arXiv:0707.4554 [hep-th]].


\bibitem{Emparan:2007en}
  R.~Emparan and A.~Maccarrone,
  ``Statistical Description of Rotating Kaluza-Klein Black Holes,''
  Phys.\ Rev.\  D {\bf 75}, 084006 (2007)
  [arXiv:hep-th/0701150].


\bibitem{Astefanesei:2006dd}
  D.~Astefanesei, K.~Goldstein, R.~P.~Jena, A.~Sen and S.~P.~Trivedi,
  ``Rotating attractors,''
  JHEP {\bf 0610}, 058 (2006)
  [arXiv:hep-th/0606244].




\bibitem{rasheed-larsen}
  D.~Rasheed,
  ``The Rotating dyonic black holes of Kaluza-Klein theory,''
  Nucl.\ Phys.\  B {\bf 454}, 379 (1995)
  [arXiv:hep-th/9505038].

  T.~Matos and C.~Mora,
  ``Stationary dilatons with arbitrary electromagnetic field,''
  Class.\ Quant.\ Grav.\  {\bf 14}, 2331 (1997)
  [arXiv:hep-th/9610013].

  F.~Larsen,
  ``Rotating Kaluza-Klein black holes,''
  Nucl.\ Phys.\  B {\bf 575}, 211 (2000)
  [arXiv:hep-th/9909102].


\bibitem{Giusto:2007tt}
  S.~Giusto, S.~F.~Ross and A.~Saxena,
  ``Non-supersymmetric microstates of the D1-D5-KK system,''
  JHEP {\bf 0712}, 065 (2007)
  [arXiv:0708.3845 [hep-th]].

\bibitem{Ceresole:2007wx}
  A.~Ceresole and G.~Dall'Agata,
  ``Flow Equations for Non-BPS Extremal Black Holes,''
  JHEP {\bf 0703}, 110 (2007)
  [arXiv:hep-th/0702088].



\bibitem{Andrianopoli:2007gt}
  L.~Andrianopoli, R.~D'Auria, E.~Orazi and M.~Trigiante,
  ``First Order Description of Black Holes in Moduli Space,''
  JHEP {\bf 0711}, 032 (2007)
  [arXiv:0706.0712 [hep-th]].



\bibitem{Ferrara:2008ap}
  S.~Ferrara, A.~Gnecchi and A.~Marrani,
  ``d=4 Attractors, Effective Horizon Radius and Fake Supergravity,''
  Phys.\ Rev.\  D {\bf 78}, 065003 (2008)
  [arXiv:0806.3196 [hep-th]].

\bibitem{Bellucci:2008sv}
  S.~Bellucci, S.~Ferrara, A.~Marrani and A.~Yeranyan,
  ``stu Black Holes Unveiled,''
  arXiv:0807.3503 [hep-th].



\bibitem{Ford:2007th}
  J.~Ford, S.~Giusto, A.~Peet and A.~Saxena,
  ``Reduction without reduction: Adding KK-monopoles to five dimensional
  stationary axisymmetric solutions,''
  Class.\ Quant.\ Grav.\  {\bf 25}, 075014 (2008)
  [arXiv:0708.3823 [hep-th]].

\bibitem{Giusto:2007fx}
  S.~Giusto and A.~Saxena,
  ``Stationary axisymmetric solutions of five dimensional gravity,''
  Class.\ Quant.\ Grav.\  {\bf 24}, 4269 (2007)
  [arXiv:0705.4484 [hep-th]].



\bibitem{Camps:2008hb}
  J.~Camps, R.~Emparan, P.~Figueras, S.~Giusto and A.~Saxena,
  ``Black Rings in Taub-NUT and D0-D6 interactions,''
  JHEP {\bf 0902}, 021 (2009)
  [arXiv:0811.2088 [hep-th]].


\bibitem{Bena:2005va}
  I.~Bena and N.~P.~Warner,
 ``Bubbling supertubes and foaming black holes,''
  Phys.\ Rev.\  D {\bf 74}, 066001 (2006)
  [arXiv:hep-th/0505166].

\bibitem{Berglund:2005vb}
  P.~Berglund, E.~G.~Gimon and T.~S.~Levi,
  ``Supergravity microstates for BPS black holes and black rings,''
  JHEP {\bf 0606}, 007 (2006)
  [arXiv:hep-th/0505167].


\bibitem{Saxena:2005uk}
  A.~Saxena, G.~Potvin, S.~Giusto and A.~W.~Peet,
  ``Smooth geometries with four charges in four dimensions,''
  JHEP {\bf 0604} (2006) 010
  [arXiv:hep-th/0509214].



\bibitem{supertube} 
  D.~Mateos and P.~K.~Townsend,
  ``Supertubes,''
  Phys.\ Rev.\ Lett.\  {\bf 87}, 011602 (2001)
  [arXiv:hep-th/0103030].
  R.~Emparan, D.~Mateos and P.~K.~Townsend,
  ``Supergravity supertubes,''
  JHEP {\bf 0107}, 011 (2001)
  [arXiv:hep-th/0106012].


\bibitem{4D5D}
  D.~Gaiotto, A.~Strominger and X.~Yin,
  ``New Connections Between 4D and 5D Black Holes,''
  JHEP {\bf 0602}, 024 (2006)
  [arXiv:hep-th/0503217].

\bibitem{Bena:2005zy}
  I.~Bena, C.~W.~Wang and N.~P.~Warner,
  ``Sliding rings and spinning holes,''
  JHEP {\bf 0605}, 075 (2006)
  [arXiv:hep-th/0512157].

\bibitem{Gaiotto:2007ag}
  D.~Gaiotto, W.~W.~Li and M.~Padi,
  ``Non-Supersymmetric Attractor Flow in Symmetric Spaces,''
  JHEP {\bf 0712}, 093 (2007)
  [arXiv:0710.1638 [hep-th]].




\end{thebibliography}
\end{document}